\documentclass[conference]{IEEEtran}

\usepackage{cite}
\usepackage{graphicx}
\usepackage{epstopdf} 
\usepackage{multirow}
\usepackage{array}
\usepackage{tabularx}
\usepackage{pbox}

\graphicspath{{./figures/}}
\usepackage{xcolor}
\usepackage[official]{eurosym}
\usepackage[nolist]{acronym}

\usepackage{algorithm}
\usepackage{algpseudocode}
\usepackage{adjustbox}

\title{Energy Efficient SDN and SDR Joint Adaptation of CPU Utilization Based on Experimental Data Analytics}

\author{
Beiran Chen, Frank Slyne and Marco Ruffini\\  
CONNECT centre, Trinity College Dublin, Ireland\\
email:
\{chenbe, slynef, marco.ruffini\}@tcd.ie
}

\begin{document}

\maketitle
\begin{abstract}

In this paper we propose a hybrid softwarized architecture
of \ac{NFV} where \ac{SDN} and \ac{SDR} components are integrated to form a cloud-based communication system. 
We analyze CPU utilization and power consumption in the OpenIreland testbed for different parameter settings and use case scenarios of this \ac{NFV} architecture.
The experiment results show different behaviour between \ac{SDN} data plane switching and \ac{SDR} in terms of CPU utilization and parallelization, which provides insights for processing aggregation and power savings when integrating them together in a cloud-based system. 
We then propose a power saving scheme with flexible CPU allocation that can reduce the overall power consumption of the system. Our results show that our proposed \ac{NFV} architecture and its power saving scheme can save up to 20\% power consumption compared to conventional scheme where \ac{SDN} and \ac{SDR} are separately deployed.

\end{abstract}

\begin{IEEEkeywords}
CPU, power saving, NFV, SDN, SDR 
\end{IEEEkeywords}

\section{Introduction}
\label{sec:introduction}

\ac{NFV} technologies are being proposed to be used broadly in next-generation networks, especially in cloud-based network architecture, e.g. \ac{MEC}, \ac{IoT}, etc \cite{vDBA, Das_JOCN}. The softwarization and virtualization of the network functions enables the opportunity for cloud technologies being embedded in the \ac{NFV}. 
\ac{SDR} and \ac{SDN} 
play important roles in next generation wireless and wired networks, as together they can enable unprecedented flexibility in capacity allocation and service differentiation \cite{SDN_SDR_magazine}. However, these are conventionally deployed separately (i.e., in different servers), which can lead to suboptimal use of resources. In this paper we investigate the power consumption of the two technologies and propose power saving methodologies, based on the coexistence of the two paradigms on the same group of CPU cores.
Our previous work~\cite{ChZhIo-IoT20} proposed a power saving strategy based on reinforcement learning by dynamically adding and reducing CPU cores for \ac{SDN} and \ac{SDR}. However, in that work the parameter settings of the active-reward pairs are assumed with normalized values. In this paper, we go deeper into the power consumption and CPU utilization by carrying out extensive testbed measurements, collecting and analyzing data of a real cloud-based system for \ac{SDN} and \ac{SDR} in the Trinity College Dublin OpenIreland testbed~\cite{open_ireland}.

There are other works related to our work. For instance, authors in~\cite{AyGaCo21} presented an experimental study of power consumption of virtualized base stations (vBSs), investigating its relationship with performance. However, they do not investigate \ac{SDN} or power saving schemes. Authors in~\cite{EiKiMo19} proposed a dynamic resource allocation scheme in Software-Defined-Radio Access Networks based on statistical evaluations. However, they focus on allocation of radio resources rather than CPU computational resources. Authors in~\cite{IsMa2020} studied several types of functional splits in the \ac{NFV} of a dual-site network in Virtualized Radio Access Networks with \ac{SDR}. The paper investigated the combined optimization of the power consumption and the mid-haul bandwidth of the functional splits. However, there was no real testbed measurements involved in the paper.
Authors in \cite{FeCeOc16} solved an optimization problem to minimize the power consumption of \ac{SDN} by Integer Linear Programming and routing constraints, however there was no testbed measurement or experiment involved either. 
Based on the investigation of related works, the contribution of our paper can be summarized as follows. 

\begin{itemize}
    \item We propose an \ac{SDN} and \ac{SDR} integrated \ac{NFV} architecture that is a cloud-based virtualization architecture.
We investigate power consumption of virtual switches and \ac{SDR} by real testbed measurement.
    \item We propose a power optimization algorithm with \ac{SDN} and \ac{SDR} integrated CPU deployment, and compare it to the case where these two functions are deployed separately.
\end{itemize}

The results of our work show different CPU utilization features for \ac{SDN} and \ac{SDR} which enables the feasibility of power savings of an integrated CPU deployment of these two important \ac{NFV} functions. Our power saving scheme can save up to 20\% power consumption compared to the approach where \ac{SDN} and \ac{SDR} are deployed separately.

\begin{figure}[h]
	\centering
	\includegraphics[width=\columnwidth]{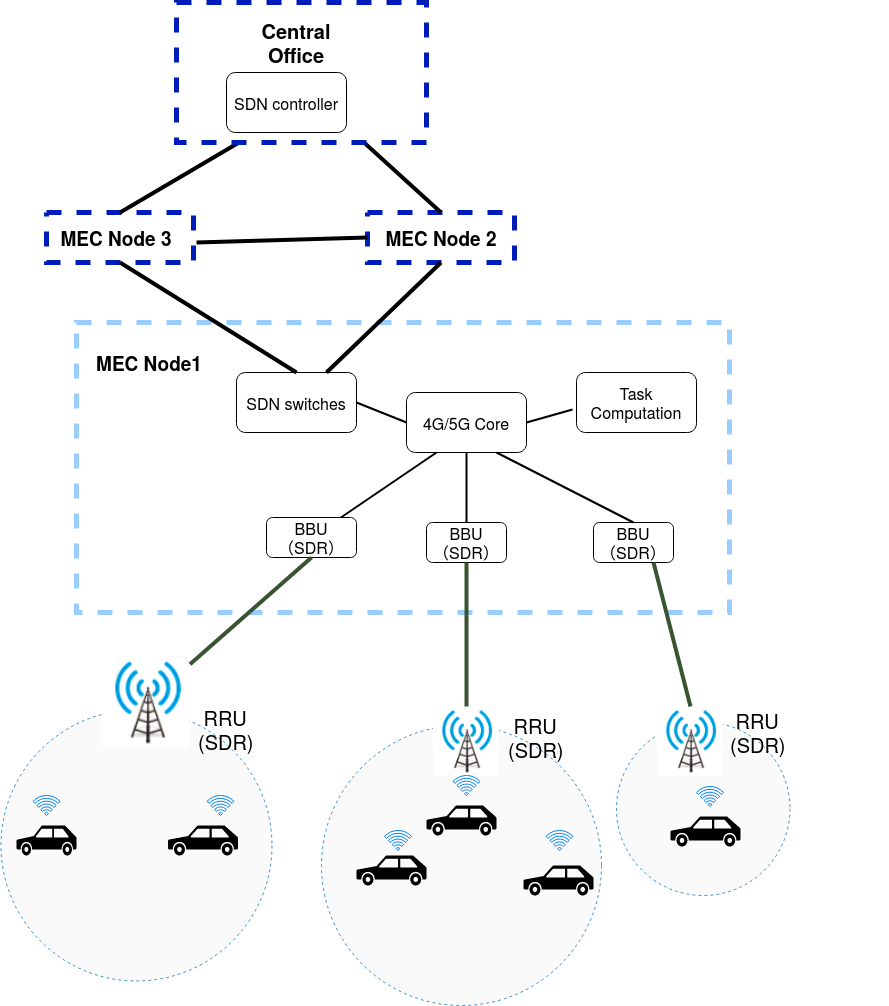}
	\caption{Proposed SDN/SDR integrated architecture with MEC as a use case example}
	\label{fig:MEC_architecture}
	\vspace{-0.1in}
\end{figure}

We propose an \ac{MEC} as an example for a use case of \ac{SDN} and \ac{SDR} integrated architecture, as shown in Figure~\ref{fig:MEC_architecture}. This architecture can be for example adopted in systems for inter-vehicle communications \cite{1543745}.
The architecture is all software-defined and network functions are virtualized. As the two major network functions for this architecture, \ac{SDN} and \ac{SDR} functions are integrated in the edge servers (i.e. \ac{MEC} nodes). Vehicular networks are illustrated at end user side. Inside the \ac{MEC} nodes, SDN switches, eNodeB \acp{BBU}, 4G/5G core function, and computational resources for \ac{MEC} tasks utilize the same group of servers. In this paper, we focus on the integration of the communication part, i.e., \ac{SDN} and \ac{SDR} components.


\section{CPU utilization in NFV}
\label{sec:CPU_utilisation}

In this section we investigate the CPU utilization for two \acp{VNF}, \ac{SDN} and \ac{SDR}, through testbed measurements and data analysis.

\subsection{CPU utilization in SDN}


In order to characterize the CPU utilization of SDN functions we have investigated multiple bandwidths options in a linear topology. The measurements focus on the data plane, i.e. the CPU consumption of Open vSwitches (OVS switches), and hosts.

\begin{figure}[h]
	\centering
	   \vspace{+0.1in}
\includegraphics[width=\columnwidth]{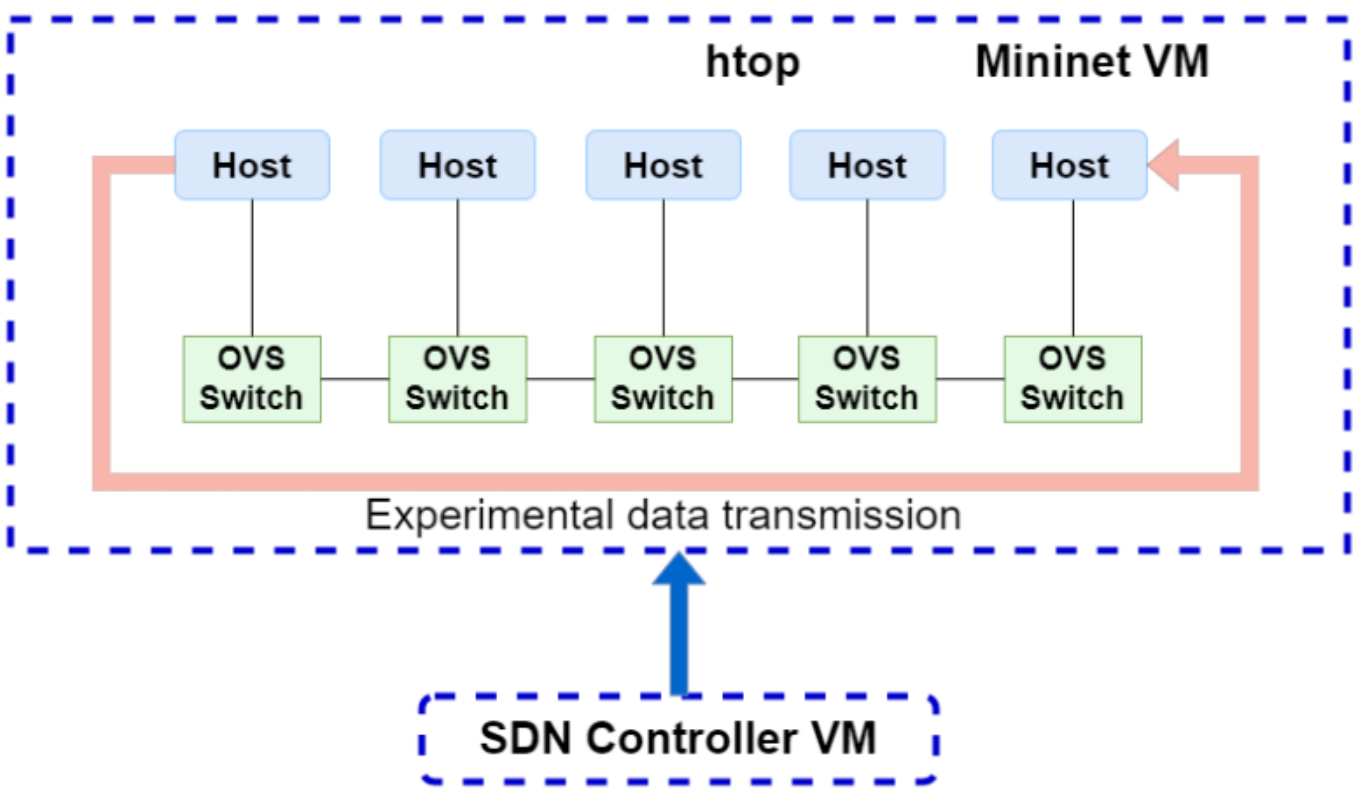}
	\caption{Experimental set up with linear topology}
	\label{fig:linear}
    \vspace{-0.1in}
\end{figure}

\begin{figure}[h]
\vspace{-0.2in}
	\centering
\includegraphics[width=\columnwidth]{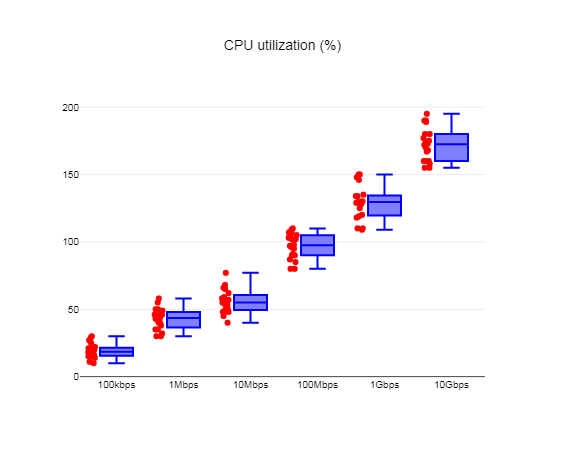}
\vspace{-20mm}
\vspace{0.15in}
	\caption{SDN CPU utilization results}
	
	\label{fig:SDN_utilization_boxplot}
\end{figure}

Fig.~\ref{fig:linear} shows our experimental set up of \ac{SDN} CPU utilization measurement. We use the Mininet simulator~\cite{mininet} for SDN and two separate virtual machines (VMs) for hosting the control plane (SDN controller) and the data plane (OVS switches and hosts), respectively. The $htop$ command is used for monitoring the CPU utilization at the data plane. We use $iperf$ UDP data transmission tool integrated in the Mininet to simulate the data transmission. We measure the CPU utilization of the OVS switches and hosts for the data plane (in the Mininet virtual machine) in different data transmission rates. In the testing scenario of linear topology, we set up 5 OVS switches with one host attached to each switch, and the data transmission is from the host on the left end to the host on the right end, i.e. the route traverse 5 OVS switches. We use the Linux application $iperf$ to generate data transmission in Mininet. We use 2 CPU cores to host the data plane and use $htop$ program in Linux to measure the CPU utilization. The CPU model we use for SDN experiment is \textit{Intel Core i7-4810MQ@2.80GHz}. $htop$ is a Linux command to monitor system resource utilization \cite{htop}. 
The results are shown in Fig. \ref{fig:SDN_utilization_boxplot}, which is a boxplot with data plots in red dots. For each data transmission rate we have measured 20 data points. As the transmission rate increases from 100kbps to 10Gbps (logarithmic scale), the CPU utilization appears to increase in a linear scale, with mean value increasing from 20\% to 175\%. Note that we are using virtual \ac{SDN} switches here implemented by virtual machines, thus the CPU utilization is affected by the CPU processing capacity. The overall CPU utilization should lower for the same data rate if we use more powerful CPUs.

\subsection{CPU utilisation in SDR}

\begin{figure}[h]
	\centering
\includegraphics[width=0.45\textwidth]{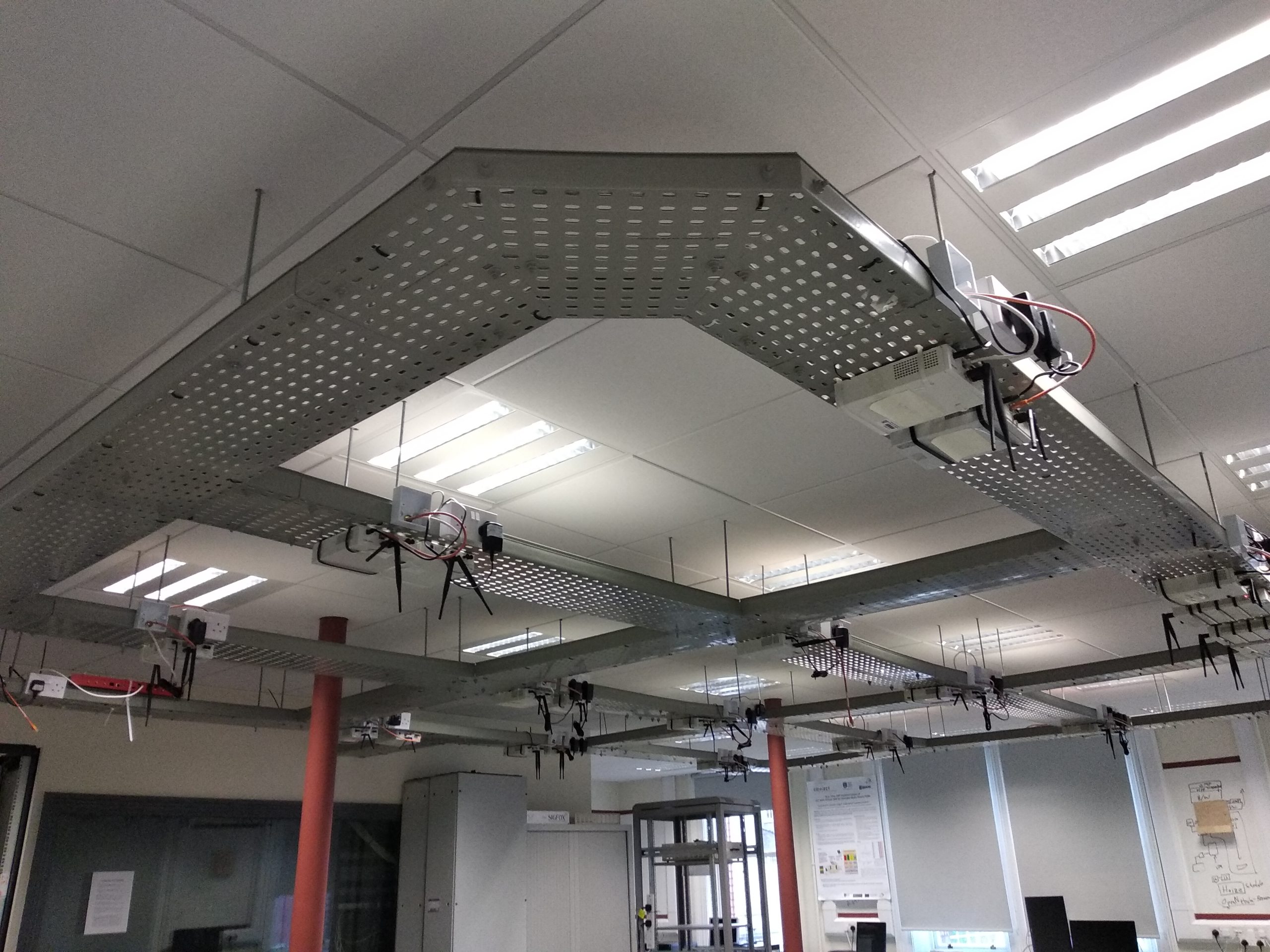}
	\caption{Testbed picture for SDR}
	\label{fig:SDR_picture}
    \vspace{0.1in}
\end{figure}

Next, we measure the CPU utilization for an SDR with different parameter settings. We use srsRAN~\cite{srsRAN} to build a software LTE base station, whose RF runs in a USRP B210. The OpenIreland testbed is located at the CONNECT centre, in Trinity College Dublin (whose SDR part is shown in Fig.~\ref{fig:SDR_picture}).

We have measured CPU utilization for scenarios involving use of different bandwidth and \acp{PRB}. Table~\ref{tab:PRB50} shows the results of 50 PRBs for the LTE eNodeB (our CPU is not powerful enough to carry out 100 PRB experiments without down sampling the wireless signals in the signal processing phase of \ac{SDR}). We have used the Linux application $iperf$ for both UDP and TCP packet transmission tests. The CPU utilization in the table is the mean value taken from 20 experimental data points. The results show that the CPU utilization increases when bandwidth settings for data transmission increases. The real throughput measured in the air is limited around 20 Mbps, and the CPU utilization is limited around 85\%-90\%. For this reason, also the packet loss for UDP and number of retries for TCP both start to increase when bandwidth setting increases beyond 20Mbps. Please note that we were using 16QAM as the modulation and coding schemes to carry out the experiments, therefore it is not able to reach the maximum throughput of 64QAM. The CPU model we use for SDR experiment is \textit{Intel NUC i7-8559U@2.70GHz}. Our CPU is not powerful enough for high modulation of 64QAM without down sampling the wireless signals in the signal processing phase of \ac{SDR}. 


Table~\ref{tab:PRB25} shows the results for the case of 25 PRBs. 16 QAM is again used in this scenario and the real throughput caps at around 10 Mbps. The CPU utilization in the table is also the mean value taken from 20 experimental data points. We can observe that the CPU utilization caps at around 60\% - 65\% percent. This is lower compared to the 20Mbps throughput using 50 PRBs, which means that in this 25 PRB scenario the full potential of the CPU processing is not reached. The data rate and CPU utilization are both constrained by the lower PRB numbers and low order of modulation and coding schemes.

\begin{table}[h]
\centering
\caption{Testbed measurement results for SDR PRB 50}
\label{tab:PRB50}
\begin{adjustbox}{width=\columnwidth, center}
\begin{tabular}{|l|lll|lll|}
\hline
PRB 50                                                                & \multicolumn{3}{c|}{TCP}                                                                                                                                                                                                                         & \multicolumn{3}{c|}{UDP}                                                                                                                                                                                                                 \\ \hline
\begin{tabular}[c]{@{}l@{}}Bandwidth \\ setting\\ (Mbps)\end{tabular} & \multicolumn{1}{l|}{\begin{tabular}[c]{@{}l@{}}Real\\ Throughput\\ (Mbps)\end{tabular}} & \multicolumn{1}{l|}{\begin{tabular}[c]{@{}l@{}}CPU \\ utilization\\ (\%)\end{tabular}} & \begin{tabular}[c]{@{}l@{}}Number\\ of\\ Retries\end{tabular} & \multicolumn{1}{l|}{\begin{tabular}[c]{@{}l@{}}Real \\ Throughput\\ (Mbps)\end{tabular}} & \multicolumn{1}{l|}{\begin{tabular}[c]{@{}l@{}}CPU \\ utilization\\ (\%)\end{tabular}} & \begin{tabular}[c]{@{}l@{}}Packet\\ Loss\end{tabular} \\ \hline
1                                                                     & \multicolumn{1}{l|}{1}                                                                  & \multicolumn{1}{l|}{28.5}                                                              & 0                                                             & \multicolumn{1}{l|}{1}                                                                  & \multicolumn{1}{l|}{29.8}                                                              & 0                                                     \\ \hline
5                                                                     & \multicolumn{1}{l|}{5}                                                                  & \multicolumn{1}{l|}{68.3}                                                              & 0                                                             & \multicolumn{1}{l|}{5}                                                                  & \multicolumn{1}{l|}{49}                                                                & 0                                                     \\ \hline
10                                                                    & \multicolumn{1}{l|}{9.8}                                                                & \multicolumn{1}{l|}{73.4}                                                              & 0                                                             & \multicolumn{1}{l|}{10}                                                                 & \multicolumn{1}{l|}{62.8}                                                              & 0                                                     \\ \hline
15                                                                    & \multicolumn{1}{l|}{14.6}                                                               & \multicolumn{1}{l|}{76.3}                                                              & 0                                                             & \multicolumn{1}{l|}{15}                                                                 & \multicolumn{1}{l|}{74.2}                                                              & 0                                                     \\ \hline
20                                                                    & \multicolumn{1}{l|}{19.3}                                                               & \multicolumn{1}{l|}{87.1}                                                              & 0                                                             & \multicolumn{1}{l|}{20}                                                                 & \multicolumn{1}{l|}{84.7}                                                              & 0                                                     \\ \hline
25                                                                    & \multicolumn{1}{l|}{21.2}                                                               & \multicolumn{1}{l|}{90.3}                                                              & 36                                                            & \multicolumn{1}{l|}{20.6}                                                               & \multicolumn{1}{l|}{86.6}                                                              & 16                                                    \\ \hline
30                                                                    & \multicolumn{1}{l|}{21.2}                                                               & \multicolumn{1}{l|}{90.3}                                                              & 58                                                            & \multicolumn{1}{l|}{20.6}                                                               & \multicolumn{1}{l|}{86.1}                                                              & 30                                                    \\ \hline
35                                                                    & \multicolumn{1}{l|}{21.1}                                                               & \multicolumn{1}{l|}{90.8}                                                              & 44                                                            & \multicolumn{1}{l|}{20.3}                                                               & \multicolumn{1}{l|}{86}                                                                & 41                                                    \\ \hline
40                                                                    & \multicolumn{1}{l|}{21.1}                                                               & \multicolumn{1}{l|}{90.9}                                                              & 30                                                            & \multicolumn{1}{l|}{20.6}                                                               & \multicolumn{1}{l|}{86.6}                                                              & 48                                                    \\ \hline
45                                                                    & \multicolumn{1}{l|}{21.1}                                                               & \multicolumn{1}{l|}{90.9}                                                              & 31                                                            & \multicolumn{1}{l|}{20.1}                                                               & \multicolumn{1}{l|}{86}                                                                & 54                                                    \\ \hline
50                                                                    & \multicolumn{1}{l|}{21.2}                                                               & \multicolumn{1}{l|}{90.4}                                                              & 40                                                            & \multicolumn{1}{l|}{20.1}                                                               & \multicolumn{1}{l|}{85.1}                                                              & 59                                                    \\ \hline
100                                                                   & \multicolumn{1}{l|}{21.2}                                                               & \multicolumn{1}{l|}{90}                                                                & 50                                                            & \multicolumn{1}{l|}{19.5}                                                               & \multicolumn{1}{l|}{86.7}                                                              & 80                                                    \\ \hline
\end{tabular}
\end{adjustbox}
\end{table}

\begin{table}[]
\centering
\caption{Testbed measurement results for SDR PRB 25}
\label{tab:PRB25}
\begin{adjustbox}{width=\columnwidth, center}
\begin{tabular}{|l|lll|lll|}
\hline
PRB 25                                                                & \multicolumn{3}{c|}{TCP}                                                                                                                                                                                                                         & \multicolumn{3}{c|}{UDP}                                                                                                                                                                                                                 \\ \hline
\begin{tabular}[c]{@{}l@{}}Bandwidth \\ setting\\ (Mbps)\end{tabular} & \multicolumn{1}{l|}{\begin{tabular}[c]{@{}l@{}}Real\\ Throughput\\ (Mbps)\end{tabular}} & \multicolumn{1}{l|}{\begin{tabular}[c]{@{}l@{}}CPU \\ utilization\\ (\%)\end{tabular}} & \begin{tabular}[c]{@{}l@{}}Number\\ of\\ Retries\end{tabular} & \multicolumn{1}{l|}{\begin{tabular}[c]{@{}l@{}}Real \\ Throughput\\ (Mbps)\end{tabular}} & \multicolumn{1}{l|}{\begin{tabular}[c]{@{}l@{}}CPU \\ utilization\\ (\%)\end{tabular}} & \begin{tabular}[c]{@{}l@{}}Packet\\ Loss\end{tabular} \\ \hline
1                                                                     & \multicolumn{1}{l|}{1}                                                                  & \multicolumn{1}{l|}{42.1}                                                              & 0                                                             & \multicolumn{1}{l|}{1}                                                                  & \multicolumn{1}{l|}{37.4}                                                              & 0                                                     \\ \hline
5                                                                     & \multicolumn{1}{l|}{4.99}                                                               & \multicolumn{1}{l|}{50.3}                                                              & 0                                                             & \multicolumn{1}{l|}{4.99}                                                               & \multicolumn{1}{l|}{51}                                                                & 0                                                     \\ \hline
10                                                                    & \multicolumn{1}{l|}{9.46}                                                               & \multicolumn{1}{l|}{61.1}                                                              & 54                                                            & \multicolumn{1}{l|}{9.13}                                                               & \multicolumn{1}{l|}{62.3}                                                              & 8.1                                                   \\ \hline
15                                                                    & \multicolumn{1}{l|}{9.46}                                                               & \multicolumn{1}{l|}{64.4}                                                              & 58                                                            & \multicolumn{1}{l|}{9.03}                                                               & \multicolumn{1}{l|}{60.1}                                                              & 39                                                    \\ \hline
20                                                                    & \multicolumn{1}{l|}{9.46}                                                               & \multicolumn{1}{l|}{63.1}                                                              & 57                                                            & \multicolumn{1}{l|}{8.92}                                                               & \multicolumn{1}{l|}{63.3}                                                              & 55                                                    \\ \hline
25                                                                    & \multicolumn{1}{l|}{9.46}                                                               & \multicolumn{1}{l|}{64.1}                                                              & 54                                                            & \multicolumn{1}{l|}{8.82}                                                               & \multicolumn{1}{l|}{63.1}                                                              & 64                                                    \\ \hline
30                                                                    & \multicolumn{1}{l|}{9.46}                                                               & \multicolumn{1}{l|}{63.7}                                                              & 58                                                            & \multicolumn{1}{l|}{8.74}                                                               & \multicolumn{1}{l|}{63.1}                                                              & 71                                                    \\ \hline
35                                                                    & \multicolumn{1}{l|}{9.46}                                                               & \multicolumn{1}{l|}{65}                                                                & 53                                                            & \multicolumn{1}{l|}{7.71}                                                               & \multicolumn{1}{l|}{63.3}                                                              & 78                                                    \\ \hline
40                                                                    & \multicolumn{1}{l|}{9.46}                                                               & \multicolumn{1}{l|}{61.7}                                                              & 59                                                            & \multicolumn{1}{l|}{7.48}                                                               & \multicolumn{1}{l|}{62.9}                                                              & 81                                                    \\ \hline
45                                                                    & \multicolumn{1}{l|}{9.47}                                                               & \multicolumn{1}{l|}{59.5}                                                              & 58                                                            & \multicolumn{1}{l|}{7.21}                                                               & \multicolumn{1}{l|}{62.1}                                                              & 84                                                    \\ \hline
50                                                                    & \multicolumn{1}{l|}{9.46}                                                               & \multicolumn{1}{l|}{63.6}                                                              & 58                                                            & \multicolumn{1}{l|}{6.54}                                                               & \multicolumn{1}{l|}{63.3}                                                              & 86                                                    \\ \hline
\end{tabular}
\end{adjustbox}
\end{table}

\section{Power Saving Methods}
\label{sec:power_savings}

In this section we investigate the power saving methods for the proposed \ac{MEC} and \ac{NFV} architecture by dynamically allocating different number of CPUs to \ac{SDN} and \ac{SDR} processes.

\subsection{Dynamic CPU allocation and process parallelisation}

In our cloud system, we propose to use the same group of CPU cores for \ac{SDN} and \ac{SDR}. We investigate the parallelization features by switching from one CPU core to two CPU cores for one process of \ac{SDN} or \ac{SDR}, using the Linux system command in $echo$ $1$ $>/sys/devices/system/cpu1/online$ and $echo$ $0$ $>/sys/devices/system/cpu1/online$ for adding/reducing CPU cores. We measure the CPU utilization for the same process running on one CPU core or on two CPU cores after one CPU is added to the virtual machine.

\subsubsection{\ac{SDN} testing}
\begin{figure}[h]
	\centering
\includegraphics[width=0.45\textwidth]{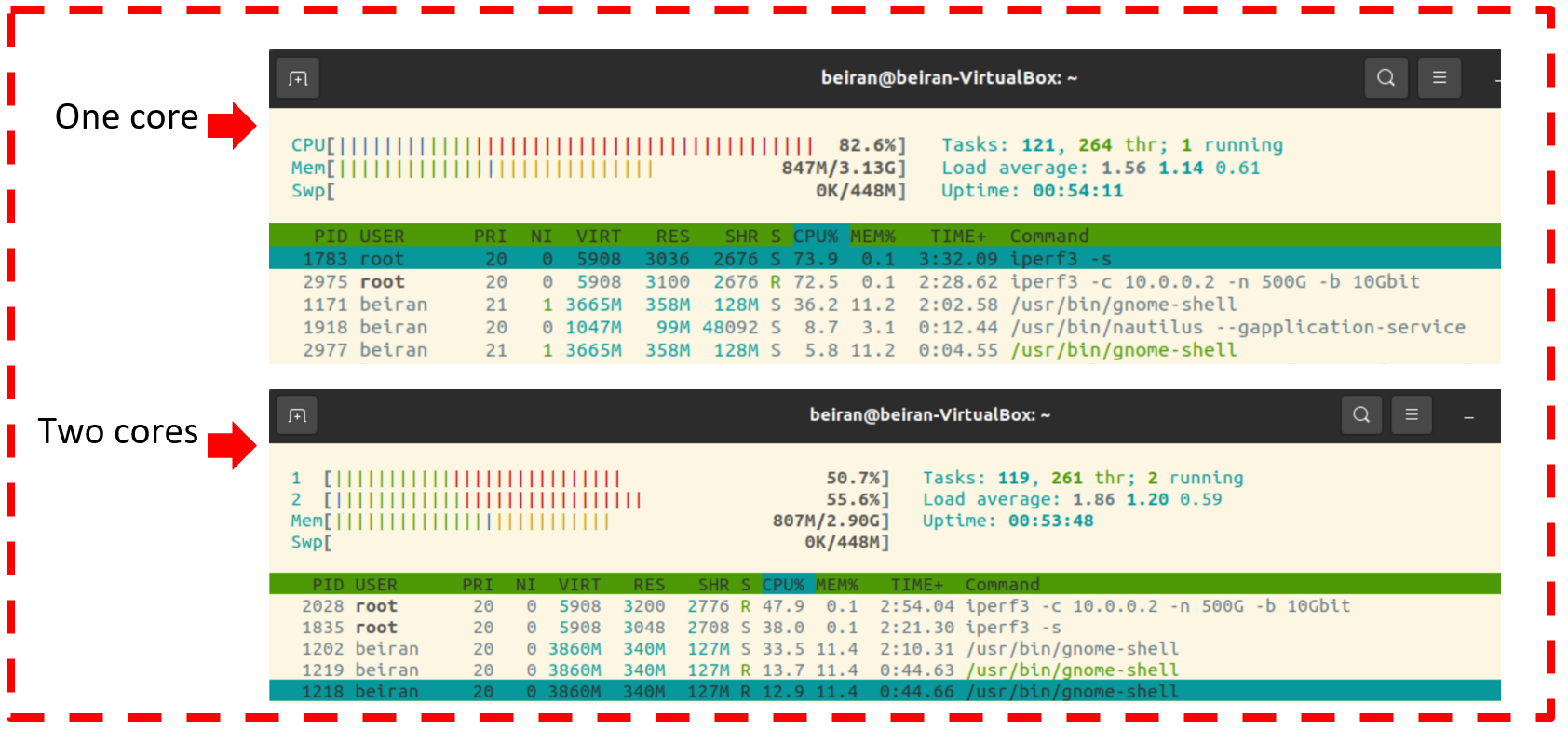}
	\caption{CPU Utilization htop measurement for SDN}
	\label{fig:htop_SDN}
    \vspace{0.1in}
\end{figure}

\begin{figure}[h]
	\centering
\includegraphics[width=0.45\textwidth]{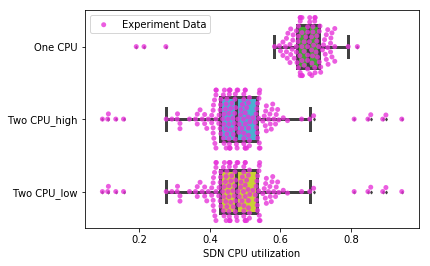}
	\caption{CPU Utilization rate for \ac{SDN}}
	\label{fig:boxplot_SDN}
    \vspace{0.1in}
\end{figure}

We use Mininet to create an SDN-controlled experimental virtual network and use $iperf$ to transmit 500GB of data, with 10Gbps bandwidth in a single link on different CPU number systems. 
We use Linux command $htop$ to measure the CPU utilization during our experiment. 
We monitor the CPU utilization rate during the transmission. When we use two CPUs to process the same \ac{SDN} task, the task appears to be parallelized on the two CPUs, with a sample $htop$ screenshot shown in Fig. \ref{fig:htop_SDN}.
We collect more samples (100 data points measured) of the CPU utilization to draw the box plots of the samples in Fig. \ref{fig:boxplot_SDN}. The results show that \ac{SDN} data plane switching processes are usually multi-threaded and can be parallelized. 

\subsubsection{\ac{SDR} testing}

\begin{figure}[h]
	\centering
\includegraphics[width=0.45\textwidth]{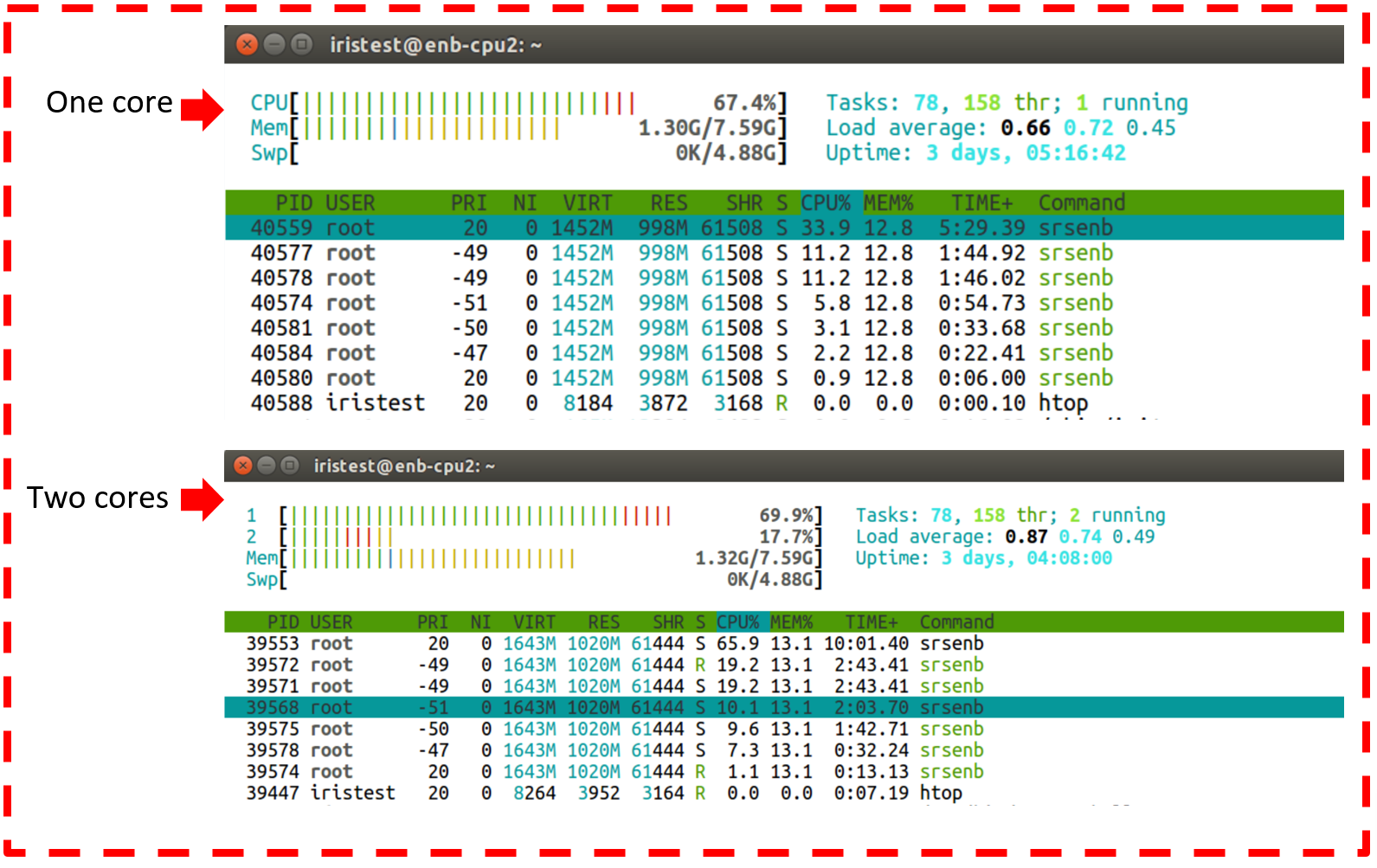}
	\caption{CPU Utilization htop measurement for SDR}
	\label{fig:htop_SDR}
    \vspace{0.1in}
\end{figure}

\begin{figure}[h]
	\centering
\includegraphics[width=0.45\textwidth]{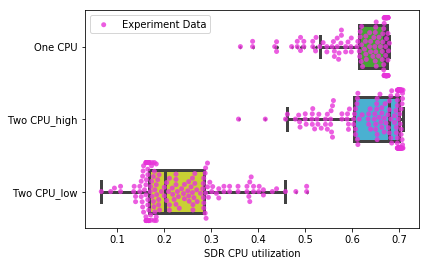}
	\caption{CPU Utilization rate for \ac{SDR}}
	\label{fig:boxplot_SDR}
    \vspace{0.1in}
\end{figure}

In this subsection, we use srsRAN to create an experimental virtual network for \ac{SDR} and use $iperf$ to transmit with 10Mbps bandwidth. We monitor the CPU utilization rate during the transmission. When we use two CPUs to process the same \ac{SDR} task, the task does not appear to be parallelized on the two CPUs, with the $htop$ sample screenshot in Fig.~\ref{fig:htop_SDR}, and more samples in box plots shown in Fig. \ref{fig:boxplot_SDR}. Among these two CPUs, the higher utilized CPU has almost the same utilization from the one CPU case and the lower utilized CPU has very low utilization. The results show that \ac{SDR} processes are usually single-thread processes that can not be parallelized.

\subsection{Shared CPU Utilization vs Separate CPU Utilization}

The algorithm we design for power savings in this paper is based on the aforementioned experimental results that the \ac{SDN} processes are multi-thread processes and easier to be parallelized, while \ac{SDR} processes are not. This gives an opportunity to optimize the total power consumption by packing the \ac{SDN} and \ac{SDR} processes in the same cluster of CPU cores. Assuming the function $P(x)$ is the function for the power consumption vs CPU utilization, the objective function is the following Equation (\ref{eqn:shared_sdn_sdr}): 

\begin{equation}
    minimise: P_{total} = \sum_{c=1}^{C}P(U(c))
    \label{eqn:shared_sdn_sdr}
\end{equation}

where $c$ is the index of the CPU core in the $C$ number of cores. $U(c)$ is the CPU utilization percentage for the CPU core $c$ and $U(c)\in [0,100]$. In this equation we don't differentiate \ac{SDN} and \ac{SDR} CPU utilization, since they share the same group of CPU cores. 

If \ac{SDN} and \ac{SDR} utilize the CPU cores separately, we have the total power consumption shown as the follow Equation(\ref{eqn:separate_sdn_sdr}):

\begin{equation}
    P_{total}=P(U_{SDN})+P(U_{SDR})
    \label{eqn:separate_sdn_sdr}
\end{equation}

In this equation, $U_{SDN}$ is the CPU utilization for \ac{SDN} and $U_{SDR}$ is the CPU utilization for \ac{SDR}. In this case the CPU utilization is decided by the processes that run separately for \ac{SDN} and \ac{SDR} without chances for optimization.

\subsection{Optimization Algorithm for Power Consumption}

We design the power consumption optimization algorithm for the \ac{SDN} and \ac{SDR} integrated system, as Algorithm \ref{alg:power_saving}.


The algorithm is an evolved version of the idea of best-fit algorithm. Since the \ac{SDR} processes are usually single thread and not capable to be parallelized, the \ac{SDR} CPU utilization comes in large chunks. On the contrary, the \ac{SDN} processes come in small chunks and can be parallelized. Therefore, the algorithm tries to fit first the large \ac{SDR} processes into the CPU cores, and then use the remaining space in CPU cores to accommodate the \ac{SDN} processes.

We have 3 usecase scenarios to test the performance of our power saving algorithm.

\begin{figure}[h]
	\centering
\includegraphics[width=0.45\textwidth]{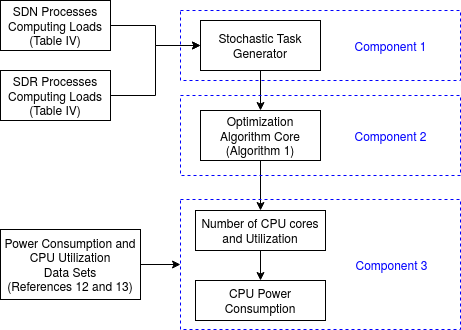}
	\caption{Simulation procedure}
	\label{fig:simulation_procedure}
\end{figure}

\begin{itemize}
    \item Usecase 1: a scenario where a large amount of \ac{SDR} bandwidth is needed but less \ac{SDN} data plane operations are needed (e.g., a local mobile network for content uploading/downloading to local servers, like mobile hotspot, Virtual Reality, Augmented Reality, etc.).
    \item Usecase 2: a scenario where little \ac{SDR} bandwidth is needed, but a large amount of \ac{SDN} data plane operations are needed, i.e., highly dynamic networking with several routing changes (e.g., mobile edge cloud and migration cases where migrations of computing tasks between edge servers are frequent, but wireless data transmission rate is low).
    \item Usecase 3: a scenario where both high \ac{SDR} bandwidth and large amount of \ac{SDN} operations are needed for both high bandwidth and high dynamicity (i.e., remote operations, mobile video streaming etc).
    
\end{itemize}
\begin{algorithm}
\caption{CPU power consumption optimization algorithm}
\label{alg:power_saving}
\begin{algorithmic}[1]
\State given $U^{i}_{SDR}, U^{j}_{SDN}$: the CPU utilization of each SDR process $i$ and SDN process $j$
\State group all SDN processes together to be $U_{SDN} = \sum_{j=1}^{J} U^{j}_{SDN}$.
\State $i \in I$ and $j \in J$
\For {all CPU core $c \in C$}
    \State Check the remaining CPU capacity of $c$: $1 - U(c)$ 
    \State Fit the largest SDR $U^{i}_{SDR}$ into $c$: $U^{i}_{SDR} \leq 1 - U(c)$
    \State Get the remaining CPU capacity: $\overline{U(c)} = 1 - U(c) - U^{i}_{SDR}$
    \State Fit the $\overline{U(c)}$ with a chunk of $U_{SDN}$ and calculate the remaining $ U_{SDN}$ by $U_{SDN} - U(c)$
    \Comment{Comment: because SDN processes can be fully parallelised according to our testbed measurement}
\EndFor

\end{algorithmic}
\end{algorithm}
Our power saving algorithm was implemented in Python and tested through extensive simulations. 
Table~\ref{tab:usecases} shows the 3 usecase settings for the process computing loads, used as input for the \ac{SDN} data plane and \ac{SDR} simulations. Each use case differs with respect to number of processes that need to be allocated to \ac{SDN} and \ac{SDR}, and CPU utilization for each process. We assume CPU utilization is randomly distributed within a given range (shown in Table~\ref{tab:usecases}) with uniform distribution. Figure \ref{fig:simulation_procedure} shows the whole procedure of our simulation, including 3 main components: the stochastic task generator (for the 3 usecases mentioned above), the power saving optimisation algorithm, and the CPU utilisation and power consumption analysis (calculating number of CPU cores and utilization, as well as power consumption). Calculation of power consumption is baseed on values derived from CPU power analysis reports \cite{spec_power_asus} and \cite{spec_power_hp}, as described below.

\begin{table}
\begin{center}
\caption{\label{tab:usecases}CPU process information for 3 usecases.}
\begin{tabular}{ |m{1.2cm} |m{1cm} |m{1.5cm} |m{1cm} |m{1.5cm} | } 
 \hline
 & \multicolumn{2}{|c|}{SDR} & \multicolumn{2}{|c|}{SDN}\\
 \hline
 Usecase \# & number of processes & CPU utilization range & number of processes & CPU utilization range\\
 \hline\hline
 Usecase 1 & 50 & [80\%, 100\%] & 30 & [10\%, 30\%]\\ 
 \hline
 Usecase 2 & 30 & [60\%, 80\%] & 50 & [30\%, 50\%]\\ 
 \hline
 Usecase 3 & 50 & [80\%, 100\%] & 50 & [30\%, 50\%]\\ 
 \hline
\end{tabular}
\end{center}
\end{table}

\begin{figure}[h]
	\centering
\includegraphics[width=0.45\textwidth]{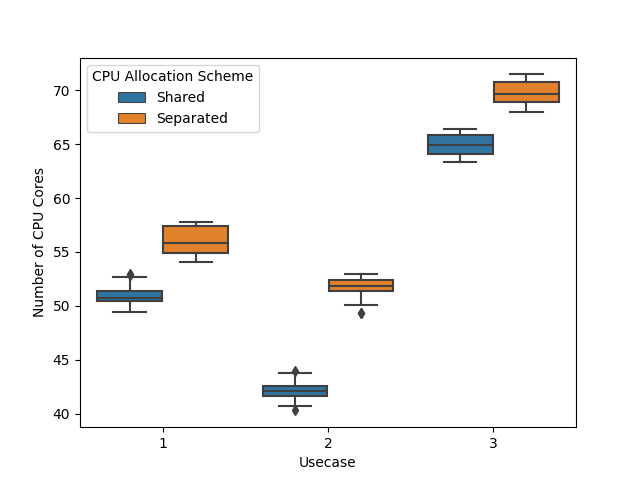}
\vspace{-0.1in}
	\caption{Number of CPU cores comparison}
	\label{fig:CPU_cores_comparison}
\end{figure}

\begin{figure}[h]
	\centering
\includegraphics[width=0.45\textwidth]{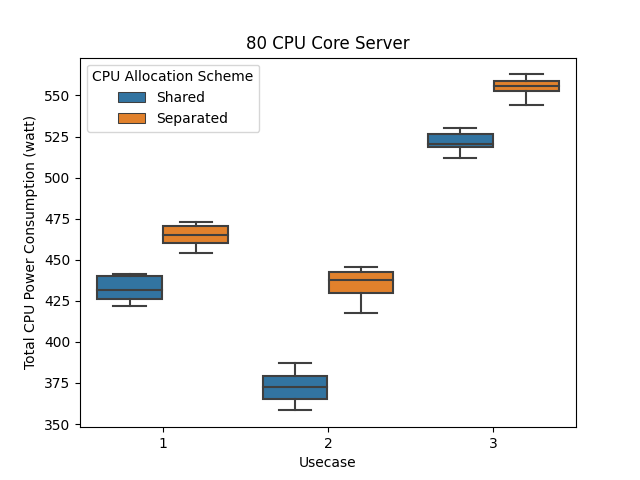}
	\caption{Power consumption comparison for 80 CPU core server}

	\label{fig:power_consumption_comparison_80}
\end{figure}

\begin{figure}[h]
	\centering
\includegraphics[width=0.45\textwidth]{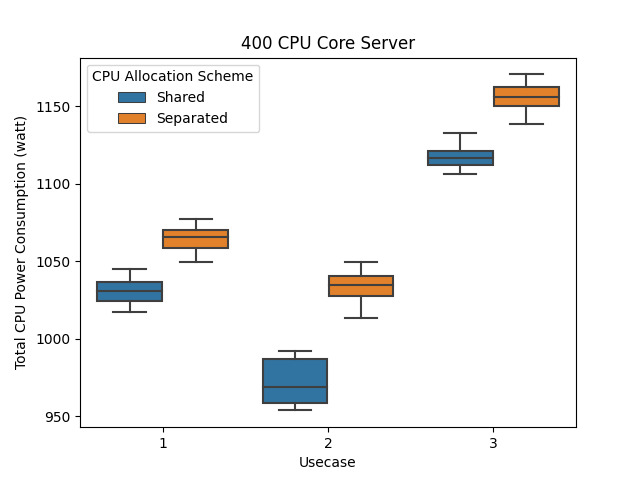}
	\caption{Power consumption comparison for 400 CPU core server}
	\label{fig:power_consumption_comparison_400}
	\vspace{-0.2in}
\end{figure}

We compare the results of our algorithm with a baseline where \ac{SDR} and \ac{SDN} processes are allocated separately in different CPU cores.
Fig.~\ref{fig:CPU_cores_comparison} shows the boxplot of total number of CPU cores used. Twenty independent tests are randomly generated for each usecase with the parameter settings in Table~\ref{tab:usecases}. 
From the results we can see that our scheme of \ac{SDN} and \ac{SDR} sharing the same group of CPU cores saves around 10\% - 20\% number of CPU cores, compared to the case of \ac{SDN} and \ac{SDR} utilizing CPU cores separately. The second use case has the most power saving percentage since in this use case the number of \ac{SDN} processes dominates, and they are easier to be parallelized, providing more room for grouping and consolidating processes in the cloud.

Fig.~\ref{fig:power_consumption_comparison_80} shows the total power consumption of these CPU cores. For this we have used power measurement data for the server model ASUSTeK Computer Inc. RS720-E10-R12, which contains 80 CPU cores, available at \cite{spec_power_asus}. Fig.~\ref{fig:CPU_cores_comparison} and Fig.~\ref{fig:power_consumption_comparison_80} shows similar trends, since the power vs CPU utilization relationship is close to linear \cite{spec_power_asus}. The power saving is around 10\% - 20\% for this server.

We also use a second power measurement dataset for a much larger server Hewlett Packard Enterprise Synergy 480 Gen10 Plus Compute Module, which contains 400 CPU cores, available at \cite{spec_power_hp}. The results are shown in Fig.~\ref{fig:power_consumption_comparison_400}. The trends are similar to the case of 80 CPU core system, but the power saving percentage is lower, which is around 5\% - 10\% percent. This is because the idle power consumption (i.e., the overhead) of this 400-core system is already 700 watt, which is much larger than the 100 watt idle power consumption (i.e., the overhead) of the 80-core system. The results in Fig.~\ref{fig:power_consumption_comparison_80} and Fig.~\ref{fig:power_consumption_comparison_400} are also boxplots generated from 20 independent tests.

There are other types of servers and CPUs that can be investigated to obtain power savings. In summary, our power saving scheme based on integrated \ac{SDN} and \ac{SDR} processing has up to 20\% power savings compared to the case where \ac{SDN} and \ac{SDR} are deployed in separate CPU cores.

\section{Conclusion}
\label{sec:conclusion}

In this paper, we have used testbed measurements with real cloud-based virtual machines to analysis power consumption of \ac{NFV} processes, in particular, \ac{SDN} and \ac{SDR} functions. We have investigated the CPU utilization and parallelization for different configurations and scenarios and developed an energy saving scheme based on the data collected from the measurements. Our power saving methods can save up to 20\% power consumption compared to the case where \ac{SDN} and \ac{SDR} are deployed separately in the system.
\begin{acronym} 

\acro{3GPP}{Third Generation Partnership Project}
\acro{BBU}{BaseBand Unit}
\acro{BLER}{Block Error Rate}
\acro{CTR}{Click Through Rate[}
\acro{D2D}{Device-to-Device}
\acro{DFT}{Discrete Fourier Transform}
\acro{DSP}{Digital Signal Processing} 
\acro{eWINE}{Elastic Wireless Network Experimentation}
\acro{FDD}{Frequency Division Duplex}
\acro{GUI}{Graphical User Interface}
\acro{IDFT}{Inverse Discrete Fourier Transform}
\acro{IoT}{Internet of Things}
\acro{ITS}{Intelligent Transportation System}
\acro{LTE}{Long Term Evolution}
\acro{MAC}{Medium Access Control}
\acro{MEC}{Multi-Access Edge Computing}
\acro{MCS}{Modulation and Coding Scheme}
\acro{MIB}{Master Information Block}
\acro{NFV}{Network Function Virtualization}
\acro{NN}{Neural Networks}
\acro{OFDM}{Orthogonal Frequency Division Multiplexing}
\acro{PDCCH}{Physical Downlink Control Channel}
\acro{PDSCH}{Physical Downlink Shared Channel}
\acro{PHY}{Physical}
\acro{PRB}{Physical Resource Block}
\acro{PSBCH}{Physical Sidelink Broadcast Channel}
\acro{PSCCH}{Physical Sidelink Control Channel}
\acro{PSDCH}{Physical Sidelink Discovery Channel}
\acro{PSS}{Primary Synchronization Signal}
\acro{PSSCH}{Physical Sidelink Shared Channel}
\acro{PSSS}{Primary Sidelink Synchronization Signal}
\acro{QAM}{Quadrature Amplitude Modulation}
\acro{QoS}{quality of service}
\acro{RNTI}{Radio Network Temporary Identifier}
\acro{RSSI}{received signal strength indicator}
\acro{SC-FDMA}{Single Carrier Frequency Division Multiple Access}
\acro{SCI}{Sidelink Control Information}
\acro{SDMA}{Space Division Multiple Access}
\acro{SDN}{Software-Defined Networking}
\acro{SDR}{Software-Defined Radio}
\acro{SNR}{Signal-to-Noise Ratio}
\acro{TDMA}{Time Division Multiple Access}
\acro{UE}{User Equipment}
\acro{UHD}{USRP Hardware Driver}
\acro{USRP}{Universal Software Radio Peripheral}
\acro{VNF}{Virtual Network Function}

\end{acronym}

\section*{Acknowledgment}

Financial support from Science Foundation Ireland (SFI) grants 17/CDA/4760 (SoftEdge), 18/RI/5721 (OpenIreland) and 13/RC/2077\_p2 (CONNECT) is gratefully  acknowledged.

\bibliographystyle{IEEEtran}
\bibliography{bibliography}
\end{document}